\documentclass[1p]{elsarticle}
\usepackage{graphics}
\usepackage{amsmath}
\begin{document}
\newcommand{\op}{\boldsymbol}
\bibliographystyle{apsrev}

\title{Quantum Key Distribution with Qubit Pairs}

\author{Mohd Asad Siddiqui}
\ead{asad@ctp-jamia.res.in}
\author{Tabish Qureshi}
\ead{tabish@ctp-jamia.res.in}
\address{Centre for Theoretical Physics, Jamia Millia Islamia, New Delhi, India.}


\begin{abstract}
We propose a new Quantum Key Distribution method in which Alice sends 
pairs of qubits to Bob, each in one of four possible states. Bob
uses one qubit to generate a secure key and the other to generate an
auxiliary key. For each pair he randomly decides which qubit to use for
which key. The auxiliary key has to be added to Bob's secure key in order
to match Alice's secure key. This scheme provides an additional layer
of security over the standard BB84 protocol.
\end{abstract}

\maketitle

Quantum cryptography \cite{qkd} is a field which has matured since the first 
protocol, BB84, given by Bennett and Brassard in 1984 \cite{bb84}.
The idea is to generate
a secret key between two remote parties, traditionally called Alice and
Bob using a quantum channel. The secret key may then be used in sending
encrypted messages through the so-called Vernam Cipher \cite{vernam} or
one-time pad. Vernam cipher has been proved to be very secure, under the
condition that one shared key is to be used only once. 

In principle, BB84 is proven to be hundred percent secure \cite{shor}.
However, when implemented in real life with non-ideal sources and detectors,
several attacks have been successfully demonstrated against commercial QKD
systems \cite{qkdbreak,qkdbreak1}.

In the traditional BB84 scheme, Alice sends a stream of single qubits to
Bob which eventually leads to the generation of a secure key. Here
we propose a new QKD scheme in which Alice sends a stream pairs of qubits
to Bob. We will show that this scheme introduces an additional feature
over the standard BB84 protocols, which should make it more difficult
to break. It should be mentioned that QKD protocols using qubit pairs have
also been proposed earlier \cite{shaari}.

To set the ball rolling, we describe the BB84 QKD protocol \cite{bb84}.
\begin{itemize}
\item Alice sends single qubits to Bob randomly in one of the following
states: $|0\rangle,~|1\rangle,~|+\rangle$ and $|-\rangle$, where
$|\pm\rangle = (|0\rangle \pm |1\rangle)/\sqrt{2}$.
\item Bob measures the incoming qubit's state by randomly choosing a 
measurement of either the x-component of the qubit or the z-component,
with equal probability. Let us say, $|0\rangle,~|1\rangle$ are eigenstates
of the z-component of the qubit, and $|\pm\rangle$ are the eigenstates of
the x-component.
\item Bob publicly tells Alice which bases he used for each qubit he received
(but, of course not the result of his measurement).
\item Alice publicly tells Bob which basis she sent each qubit in. 
\item Alice and Bob keep only the data from those measurements for which their
bases are the same, discarding all the rest.
\item This data is interpreted as a binary sequence according to the coding
scheme $|0\rangle_x$ = 0, $|1\rangle_x$ = 1, $|+\rangle_z$ = 1,
$|-\rangle_z$ = 0.
\item Alice announces the results of a small subset of her measurements. 
Bob checks if he has identical results. Any discrepency here indicates
a possible evesdropping attempt.
\item If there is no discrepancy, the rest of the binary sequence is treated
as the new key, and is identical for both Alice and Bob.
\end{itemize}

In our new scheme, Alice sends pairs of qubits to Bob, randomly chosen to be
in one of the following states,
\begin{eqnarray}
&&|1\rangle_1|1\rangle_2,\nonumber\\
&&|0\rangle_1|0\rangle_2,\nonumber\\
&&\frac{1}{\sqrt{2}}(|+\rangle_1|+\rangle_2+|-\rangle_1|-\rangle_2)\nonumber\\
&&\frac{1}{\sqrt{2}}(|+\rangle_1|-\rangle_2+|-\rangle_1|+\rangle_2),
\end{eqnarray}
where $|\pm\rangle_i = (|1\rangle_i \pm |0\rangle_i)/\sqrt{2}$.
For Alice, the key bits associated with each state she sends are as follows:
\begin{eqnarray}
|1\rangle_1|1\rangle_2&\rightarrow& 1 ,\nonumber\\
|0\rangle_1|0\rangle_2&\rightarrow& 0 ,\nonumber\\
\frac{1}{\sqrt{2}}(|+\rangle_1|+\rangle_2+|-\rangle_1|-\rangle_2)
&\rightarrow& 0 ,\nonumber\\
\frac{1}{\sqrt{2}}(|+\rangle_1|-\rangle_2+|-\rangle_1|+\rangle_2)&\rightarrow& 1.
\end{eqnarray}
In addition, she calls the first two states as z-basis, and the other two
as x-basis.

Bob uses one qubit for his secure key and one for the auxiliary key.
For generating the secure key, he randomly measures the x-component
or the z-component of the qubit. For the auxiliary key he measures only
the x-component.
For his secure key measurement results, he uses the following convention
for key bit values:
$|1\rangle \rightarrow 1$, $|0\rangle \rightarrow 0$,
$|+\rangle \rightarrow 1$, $|-\rangle \rightarrow 0$.
For his auxiliary key measurement results, he uses the following convention
for key bit values:
$|+\rangle \rightarrow 1$, $|-\rangle \rightarrow 0$, if he measured 
x-component for the secure key;
$|+\rangle \rightarrow 0$, $|-\rangle \rightarrow 0$, if he measured 
z-component for the secure key.

Alice and Bob announce their bases for secure key publicly, and discard
those qubits for which their bases do not match.
Various measurement results for rest of the cases where the bases agree,
will be correlated in the following fashion.
If Alice sends z-basis, the secure key bits of Alice and Bob will be
identical. If Alice sends x-basis, the secure key bits of Alice and
Bob will be identical if the auxiliary key bit is 0; the secure key bits of
Alice and Bob will be different if the auxiliary key bit is 1.

Thus,for a sequence of qubit pairs sent, Alice and Bob's secure key bits
will be identical if the corresponding auxiliary bit is 0. Alice and
Bob's secure key bits will not match for cases where the auxiliary bit is 1.
So, Bob just needs to add his final auxiliary key to his secure key
bit by bit, modulo 2. This way, his auxiliary bit 1 added to his 
non-matching secure key bit will make it identical to Alice's bit. 
The result of a typical sequence of qubit pairs is shown in Table \ref{demo}.

An evesdropper trying to figure out the shared key in between, will have
his task made difficult in more ways than it is in BB84. For correctly
figuring out the auxiliary key, the evesdropper has to know which of the two
qubits was used for the auxiliary key for each and every pair, an impossible
task. If we assume that the evesdropper is able to correctly guess which
of the two qubits is used for the auxiliary key, he can simply make an
x-basis measurement on those qubits, and generate the correct
auxiliary key without Bob's knowledge. However, even in this near impossible
circumstance, the rest of the communication still remains as the 
standard BB84.

However, there is subtlety involved here. If the eavesdropper attacks pairs
of qubits, then she gains more information attacking this protocol than BB84,
because the second qubit adds a certain amount of redundancy. In order to
deal with such an attack, we introduce an addition step in the protocol.
Alice generates pairs of entangled particles, but before sending them to
Bob, she rearranges the order of particles in one of the channels using
a predecided ordering-key shared with Bob. Now when she send out pairs of particles,
the two particles in a pair are not in general correlated, but one particle
of a pair is correlated with the second particle of another pair. Eve
cannot know which pairs are correlated. After Bob receives all the pairs,
he reorders particles in the relevant channel according the the ordering-key.
He then proceeds with the rest of the protocol as before. This reordering 
scheme is in the spirit of similar schemes proposed earlier \cite{deng}.
With the reordering in place, the security of this protocol should be 
{\em at least} equal to that of BB84.

\begin{table}
\noindent\begin{tabular}{|c|ccccccccccccccc|}
\hline
Basis & x & z & x & x & z & x & z & z & x & z & z & x & z & x & x \\
Alice key bit& 0 & 1 & 1 & 0 & 1 & 0 & 0 & 0 & 1 & 1 & 1 & 0 & 0 & 0 & 1\\
Bob secure state&$-$& 1 &$-$&$-$& 1 &$-$& 0 & 0 &$-$& 1 & 1 &$-$& 0 & + &$-$\\
Bob secure key bit& 0 & 1 & 0 & 0 & 1 & 0 & 0 & 0 & 0 & 1 & 1 & 0 & 0 & 1 & 0\\
Bob auxiliary state&$-$& + & + &$-$&$-$&$-$& + &$-$& + &$-$& + &$-$&$-$& + & +\\
Bob auxiliary bit  & 0 & 0 & 1 & 0 & 0 & 0 & 0 & 0 & 1 & 0 & 0 & 0 & 0 & 1 & 1\\
Bob secure + auxil& 0 & 1 & 1 & 0 & 1 & 0 & 0 & 0 & 1 & 1 & 1 & 0 & 0 & 0 & 1\\
\hline
\end{tabular}
\caption{Result of a typical sequence of 15 qubits sent by Alice to Bob,
where the measurement basis of Alice and Bob matches. When the auxiliary
key is added to Bob's secure key, it matches with Alice's key.
}
\label{demo}
\end{table}

In conclusion, we have introduced a new QKD scheme in which Alice sends
pairs of qubits to Bob in four possible states. Bob randomly chooses 
one of the pair for his secure key and one for the auxiliary key. 
The auxiliary key has to be added to Bob's secure key in order to get the
correct shared key. An evesdropper's job is made more difficult as he has
to correctly predict which particle Bob is going to use for secure key
and which one for auxiliary key, for every pair used in the communication.
Even if the evesdropper manages to acheive this near impossible feat,
he is still left with the job of cracking the BB84 security. In addition,
the scheme is modified using a reordering of particles to deal with two
qubit attacks.  We believe
this scheme should be implementable in practice.

\section*{Acknowledgments}
Asad Siddiqui thanks the University Grants Commission for
financial support. The authors thank an anonymous referee for constructive
suggestions.

\end{document}